\documentclass[pageno]{jpaper}

\newcommand\headercell[1]{%
   \smash[b]{\begin{tabular}[t]{@{}c@{}} #1 \end{tabular}}}
\usepackage{tikz}
\newcommand*\circled[1]{\tikz[baseline=(char.base)]{
            \node[shape=circle,draw,inner sep=2pt] (char) {#1};}}

\usepackage[normalem]{ulem}
\usepackage{algorithm}
\usepackage{algorithmic}
\usepackage{amsmath,amssymb,amsfonts}
\usepackage{multirow}
\usepackage{xcolor}
\usepackage{booktabs}
\usepackage{graphicx}
\usepackage{textcomp}

\begin{document}

\title{RIPencapsulation: Defeating IP Encapsulation on TI MSP Devices}

\author{Prakhar Sah\\
    sprakhar@vt.edu\\
    Virginia Tech
    \and
    Matthew Hicks\\
    mdhicks2@vt.edu\\
    Virginia Tech}

\date{}
\maketitle

\thispagestyle{empty}

\begin{abstract}

Internet of Things (IoT) devices sit at the intersection of unwieldy software complexity and unprecedented attacker access.
This unique position comes with a daunting security challenge: how can I protect both proprietary code and confidential data on a device that the attacker has unfettered access to?
Trusted Execution Environments (TEEs) promise to solve this challenge through hardware-based separation of trusted and untrusted computation and data.
While TEEs do an adequate job of protecting secrets on desktop-class devices, we reveal that trade-offs made in one of the most widely-used commercial IoT devices undermine their TEE's security.

This paper uncovers two fundamental weaknesses in IP Encapsulation (IPE), the TEE deployed by Texas Instruments for MSP430 and MSP432 devices.
We observe that lack of call site enforcement and residual state after unexpected TEE exits enable an attacker to reveal all proprietary code and secret data within the IPE.
We design and implement an attack called RIPencapsulation, which systematically executes portions of code within the IPE and uses the partial state revealed through the register file to exfiltrate secret data and to identify gadget instructions.
The attack then uses gadget instructions to reveal all proprietary code within the IPE.
Our evaluation with commodity devices and a production compiler and settings shows that---even after following all manufacturer-recommended secure coding practices---RIPencapsultaion reveals, within minutes, both the code and keys from third-party cryptographic implementations protected by the IPE.

\end{abstract}

\section{Introduction}

The global IoT industry is projected to become a trillion-dollar industry by 2027~\cite{noauthor_undated-yv}.
IoT devices are widely deployed in both safety- and mission-critical roles in government, healthcare, transportation, manufacturing, defense, and telecommunications industries.
As such, these devices are a treasure trove of sensitive information.
Data security concerns are a major obstacle to the growth of the IoT sensor market as data breaches continue to rise with the advancement of technology.
In addition to data security, the growing software complexity of IoT devices and proliferation of Artificial Intelligence mandate code security, i.e., the protection of proprietary algorithms and models.
Failing to protect proprietary code and secret data puts both consumers and companies at risk.

``Trying to design information security solutions without due consideration of the complex human nature may prove to be an Achilles heel"~\cite{akhunzada2015man}.
Cryptographic algorithms like AES and RSA provide confidentiality of data, but the key still ends up in device memory.
This leaves keys vulnerable to exfiltration by an attacker with physical access.
Unfortunately, physical access is the common case for IoT devices. 
To address the threat of attackers with such access, device manufacturers provide a Trusted Execution Environment (TEE).
TEEs bifurcate hardware (either physically or virtually) into security domains, where code and data in the high-security domain are protected from the low-security domain.
For IoT-class devices, Texas Instruments provides a TEE called IP Encapsulation (IPE), which physically partitions device memory into a protected region and an unprotected region.

Texas Instruments is the world's second-largest manufacturer of microcontrollers~\cite{Utmel2022-lr}.
Their MSP family of devices is one of the most widely deployed class of microcontrollers ~\cite{eetimes2019, embeddedsurvey2023}, with over 2000 devices in its portfolio~\cite{noauthor_undated-bq}.
MSP430s are 16-bit industrial-grade microcontrollers with low power consumption at a low cost. 
The MSP432 line of microcontrollers extends the capabilities of the MSP430 with a 32-bit, ARM-based architecture.
Both series of devices have TEE support in the form of IPE.
These factors make the MSP series of microcontrollers the most widely deployed with a TEE.

While the community continues to probe the security of TEEs provided by higher-end devices, the security of IoT-class TEEs remains unexplored.
This paper fills that gap by analyzing the security of TI's IPE.
IPE protects code and data within the IPE zone from all non-IPE zone read and write accesses~\cite{msp430fr58xx2017msp430fr6xx, msp430ipe, msp432p4xx2016}.
IPE is enforced by the Memory Protection Unit (MPU) for MSP430 and the System Controller (SYSCTL) module for MSP432, which restricts direct external IPE zone accesses by checking the origin of memory accesses.
According to the TI documentation, the MPU/SYSTCL also restricts all JTAG/DMA accesses inside the IPE zone.
Thus, only code stored inside the IPE memory can access the data in the IPE zone, making it ideal for storing secret data (e.g., keys) and proprietary code (e.g., AI models).

Our exploration of IPE security reveals that---despite following the TI-recommended secure programming practices---the state-of-art compilers with and without their optimizations produce assemblies that leak information via the unprotected register file, rendering IP encapsulation insecure.
We observe that IPE has two weaknesses that undermine their security: (1) they allow all code outside the IPE zone to branch to arbitrary instructions within the IPE zone and (2) when execution leaves the IPE zone unexpectedly (e.g., via an interrupt) the contents of the register file remain.
We construct an attack leveraging these weakness called RIPencapsulation, which exfiltrates all code and data protected by the IPE---within minutes.
RIPencapsulation combines interrupt-based control flow attack patterns, with data-oriented attack patterns, and side-channel attack patterns to break IPE.

We implement and evaluate RIPencapsulation on TI MSP430 and MSP432 devices.
Our evaluation shows that RIPencapsulation reveals the entire contents of the IPE zone with third-party implementations of AES, SHA256, and RSA, using a variety of optimization settings on a production compiler.
RIPencapsulation exfiltrates all keys and code from both devices, automatically, within minutes---even when following all of TI's secure coding practices.

In summary, this paper makes the following technical contributions:
\begin{itemize}
    \item \textbf{Create a side channel:} We design and implement an interrupt-based side channel that reveals the state within the IPE zone of TI MSP devices (\S\ref{design_main}, \S\ref{implement_main}, \S\ref{ripe_msp432}).
    \item \textbf{Reconstruct firmware:} We use partial state, timing, and size information to reconstruct the IP-encapsulated assembly (\S\ref{design_phase2}, \S\ref{implement_phase2}).
    \item \textbf{Demonstrate generality:} We show the problem is pervasive across TI MSP devices, memory types, cryptographic implementations, and compiler optimizations (\S\ref{eval_main}, \S\ref{msp432_res}).
    \item \textbf{Discuss mitigation:} We qualitatively analyze various mitigation techniques with respect to their impact on an IoT device (\S\ref{mitigation}).
\end{itemize}

\paragraph{Responsible disclosure.} 
We have disclosed the interrupt-based data-oriented attack, RIPencapsulation, and the threat it poses to the security guarantees of IP Encapsulation provided on the MSP430 to Texas Instruments.

\section{Background}

In this section, we discuss Trusted Execution Environments as well as concepts like code reuse attacks, call site verification, and context clearing.

\subsection{Trusted Execution Environment}

\begin{figure}[htbp]
\centerline{\includegraphics[width=8.5cm]{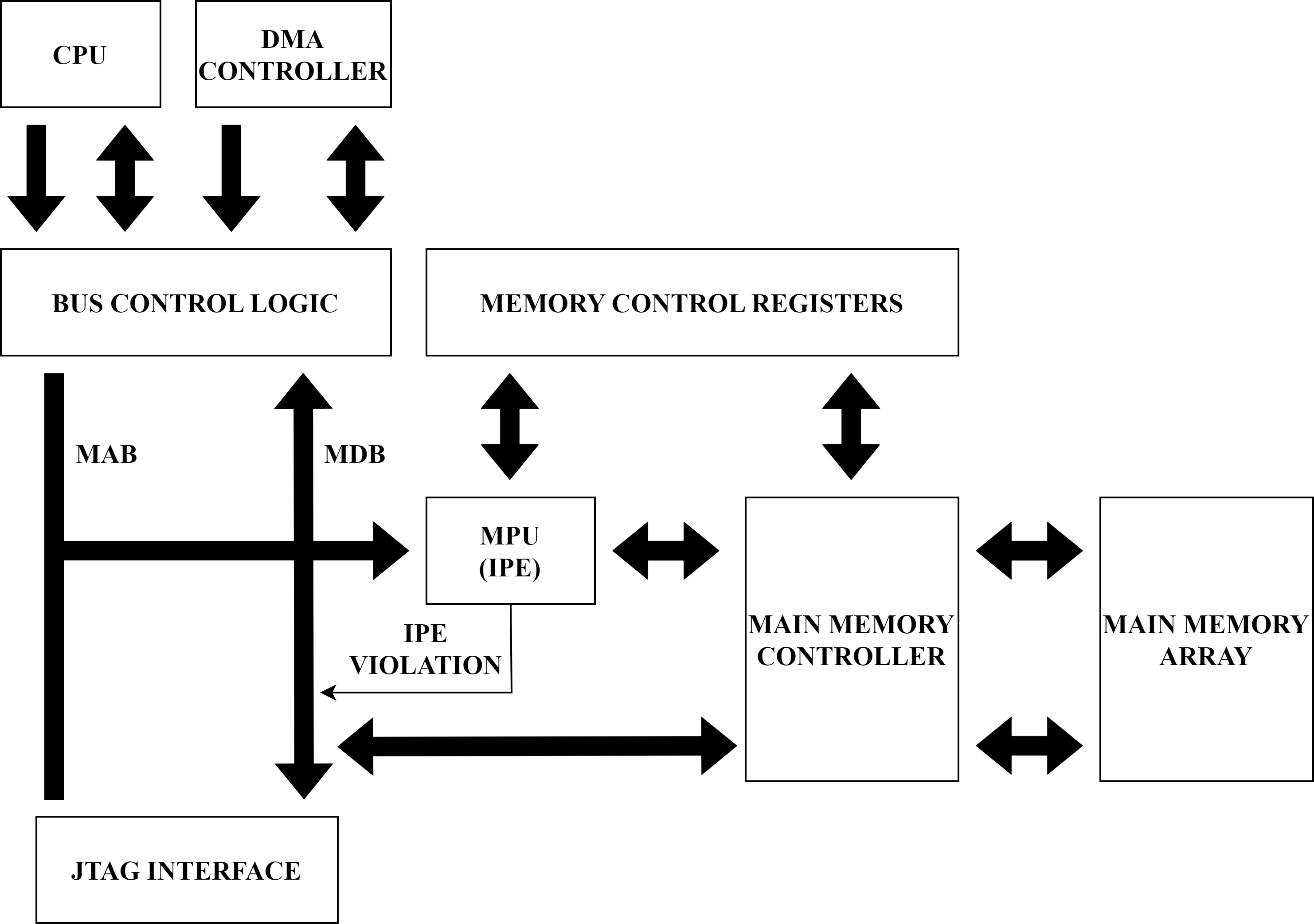}}
\caption{Functional block diagram illustrating how the MPU enforces IP Encapsulation on MSP430 devices.}
\label{MPU_Block_Diagram}
\end{figure}

A Trusted Execution Environment (TEE) is a secure processing area that provides isolated execution and secure storage of code and data inside a tamper-resistant module.
TEEs include process isolation, the integrity of applications running inside the TEE, and the confidentiality of data associated with it.
TEEs achieve this by using a memory protection mechanism restricting access to the security-critical software module~\cite{sabt2015trusted}.

Intel Software Guard Extensions (SGX) and AMD Secure Encrypted Virtualization (SEV) are multi-tenant systems running with operating systems and full-fledged commodity desktop/server-class systems.
They use hardware memory-based encryption to isolate sensitive code and data.
This type of deployment scenario is not available in low-end microcontroller devices like the MSPs due to the hardware overhead of the encryption engine and the run time overhead of encryption in an environment with single-cycle main memory access.

ARM TrustZone is a microcontroller-focused TEE that splits the physical memory into secure and normal regions, with specifically assigned computational units like Secure Attribution Unit (SAU), ensuring safe context switches between secure and normal processes, or lightweight remote attestation schemes like SMART~\cite{eldefrawy2012smart} and VRASED~\cite{nunes2019vrased}, which provide user trust with minimal HW/SW modifications.
SANCUS~\cite{noorman2013sancus} is another secure architecture that provides isolated execution and privacy of code and data in low-level networked embedded systems.

The only TEE for the low-end devices common to large-scale IoT deployments is TI's  Intellectual Property Encapsulation (IPE).
IPE protects the encapsulated memory region from all direct non-IPE accesses, whether it be from on-chip execution or off-chip via the debugger, i.e., only program code executed from the IPE region itself has access privileges to IPE code and data.
Figure~\ref{MPU_Block_Diagram} illustrates the IPE implementation for MSP430 devices: the Memory Protection Unit (MPU) verifies IPE region accesses by snooping the Memory Address Bus (MAB) and the program counter to check whether the access request is made by code in the IPE zone.
Any unauthorized access to the IPE zone causes the MPU to drive the Memory Data Bus (MDB) with \texttt{0x3FFF}.
To execute code stored inside the IPE segment, the program must call functions within the IPE zone.
The MSP432's IPE mechanism is slightly different technically, but works in a similar manner and has the same flaws as the MSP430 IPE (see \S\ref{design_main}).

No form of code protection is perfect, and TEE is no exception. There are a couple of attacks in the literature that illustrate ways to undermine such secure enclaves. CLKscrew~\cite{tang2017clkscrew} exploits software-exposed energy management mechanisms to introduce faults in the secure part of the memory, exfiltrating IP from TrustZone. Volt Boot~\cite{mahmod2022sram} is another attack that demonstrates the vulnerability of on-chip volatile memories due to the physical separation common to modern system-on-chip power distribution networks. Recent research reveals vulnerabilities in TEEs like ARM TrustZone~\cite{cerdeira2020sok} and AMD SEV~\cite{li2021cipherleaks, li2022systematic}, calling into question the reliability of these protection mechanisms. Our work adds to the TEE attack literature with an attack on TI's MSP IPE.

\subsection{Code-reuse Attacks}

Attackers turn to code reuse attacks when injecting their own code is prevented.
Instead of injecting new code, the attacker constructs malicious functionality by chaining existing code snippets, called ``gadgets'' found in the target program.
These gadgets are short sequences of instructions, typically ending with a ``return'' instruction.
By crafting a chain of gadgets and manipulating the program's control flow, the attacker is able to execute arbitrary commands.
Many techniques exist in the literature that utilize various aspects of the memory control plane~\cite{shacham2007geometry, bletsch2011jump, checkoway2010return, bosman2014framing, bittau2014hacking} and data plane~\cite{chen2005non, hu2016data} to create Turing-complete gadgets for malicious purposes.
We take inspiration from such techniques, replacing a \texttt{ret} with a carefully crafted interrupt and leveraging data-oriented attack principles.
 
\subsection{Context Switches and Call Site Verification}

Context switching refers to the process of transitioning the execution from the regular, untrusted zone to the secure, trusted zone and vice versa. 
When a context switch occurs, the current state of the regular execution environment, including registers, memory contents, and program counter, is saved, and the system transitions between zones. 
Secure context switching in TEEs is crucial for ensuring the isolation of trusted code and data from the rest of the system as any data that remains from secure zone execution serves as a side channel to untrusted execution.

In addition to context clearing, a TEE must enforce call site verification to ensure that untrusted code interfaces with trusted code in acceptable ways. 
When a program interacts with the TEE, it makes calls to functions or procedures within the enclave.
These calls are known as "call sites." 
Call site verification verifies that the callee is a function allowed to be called from untrusted code. 
This prevents attackers from hijacking the control flow and executing malicious code within the enclave.
We show that call site verification is a necessary component of any TEE, otherwise, code reuse attacks are possible.



\section{Threat Model}

The defender’s code is bug-free.
The defender follows TI's secure coding practices: (1) they clear IPE state on exit and (2) they disable interrupts upon IPE entry.
The defender uses commodity cryptographic algorithms and compilers/settings.

We assume the attacker has the capability to run any code in the untrusted world, this can either come from them having physical access~\cite{akhunzada2015man} or a buffer overflow in the non-IPE code.
This means the attacker can configure the timer and interrupt service routines.
We assume the attacker's software has a way to communicate with the attacker;
The attacker has two primary objectives: (1) exfiltrate the key of the IP-encapsulated cryptographic implementation and (2) exfiltrate all the IPE-protected code and data.
Note that this level of access is in-keeping with how TI expects users to interface with the IPE, i.e., the IPE is a place for protected, third-party code that untrusted user software can use as a library.
\section{RIPencapsulation Design}
\label{design_main}

\begin{figure}[!t]
\centerline{\includegraphics[width=8.5cm]{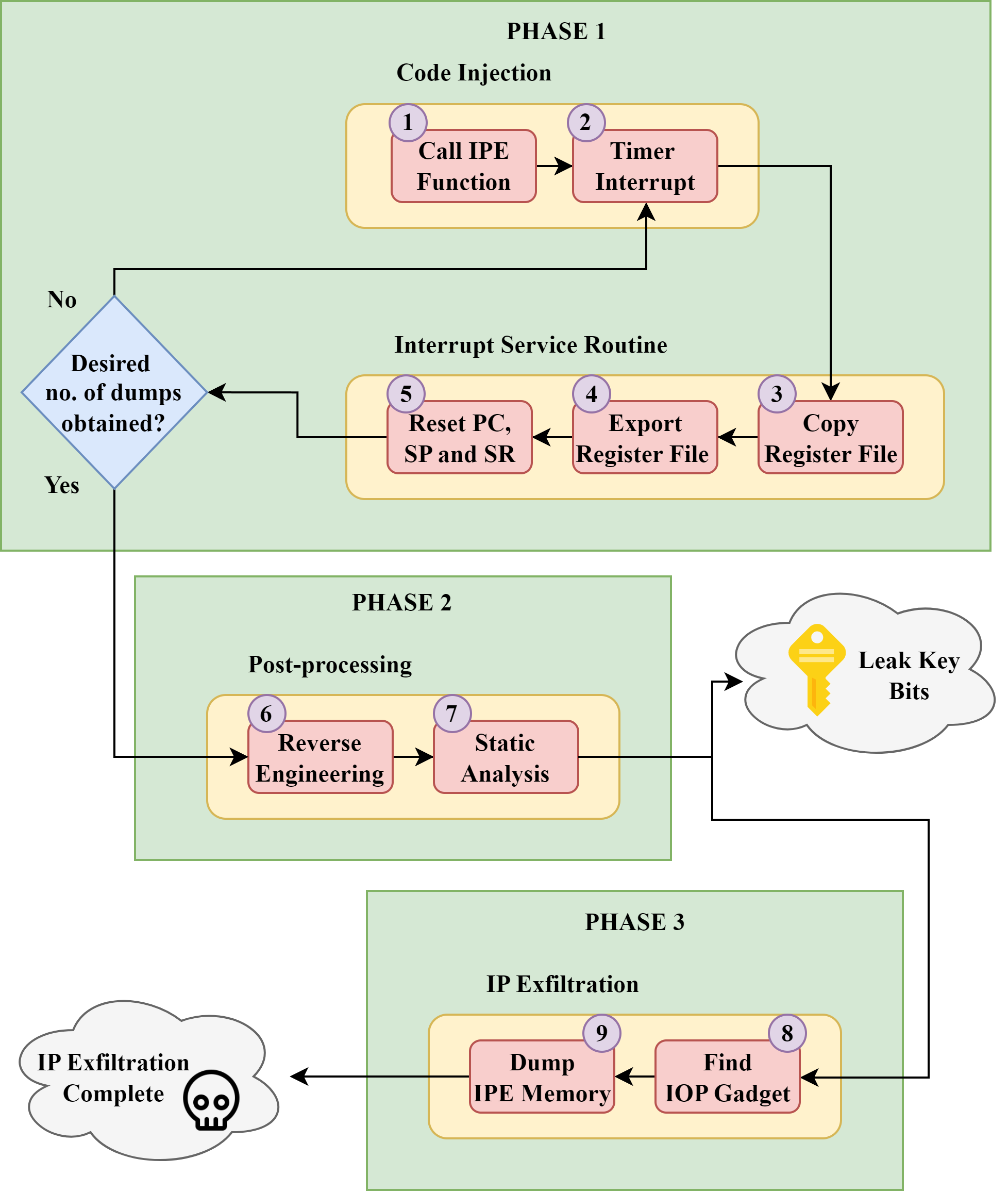}}
\caption{High-level IPE attack flow.}
\label{Design_Overview}
\end{figure}

RIPencapsulation is an interrupt-based side-channel memory exfiltration attack on TI MSP IP-Encapsulated (IPE) memory.
As Figure~\ref{Design_Overview} illustrates, the attack consists of three phases.
After the first two phases of the attack, we are able to reverse engineer 80\% of the IPE-protected code.
If the attacker's goal is only to get the keys from commodity cryptographic implementations, this generally suffices, as we prove in \S\ref{eval_main}.
However, completing all three phases enables the attacker to exfiltrate 100\% of IPE-protected memory.

\subsection{Creating a side channel (Phase 1)}
\label{design_phase1}

The fundamental weakness in the design of IPE is that it does not protect the register file when context switches to outside the IPE zone.
This first phase exploits this weakness as a side channel that reveals the internal state of the IPE zone.
The naive approach is to single-step IPE code using a debugger, but the MPU/SYSCTL monitors the memory address bus and prevents JTAG or DMA from accessing the IPE-protected memory zone.
This prevents breakpoints from working inside the IPE zone.
While we observe that it is still possible to send \texttt{HALT} signals while in the IPE zone, it only provides timing granularity of 250 $ms$;
the ideal result is single-instruction precision.

The ability to process signals indicates that, in general, interrupts are processed while executing in the IPE zone.
Thus, we can leverage a more precise interrupt source for single-instruction precision: the timer.
Timer module interrupts have the capability to interrupt a program at single-cycle granularity (i.e., sub-instruction level).

Figure~\ref{Design_Overview} shows the high-level flow of the RIPencapsulation attack.
In \circled{1}, attacker-controlled code sets up the timer module to interrupt the processor at a counter value of one clock cycle into the IPE code.
Then the attacker code jumps directly to the instruction after the instructions which disables external interrupts.\footnote{TI's secure coding practice recommends disabling interrupts while executing in the IPE zone. We leverage the lack of call site verification to bypass any interrupt disabling code. It is also possible to use external interrupts or even a power reset in place of timer interrupts, should the attack scenario require it (i.e., disabling interrupts is ineffective)}
In \circled{2}, the victim IPE code executes until the timer module interrupts it and passes control to the attacker's ISR.
It is important to note that on an interrupt, any currently executing instruction is completed, and the PC of the next instruction, along with the SR, is pushed onto the stack.
The context of the IPE code at the time of the interrupt remains intact.

Inside the ISR, \circled{3} we copy the register states to an attacker-controlled memory region.
This includes all the general-purpose registers as well as the stack pointer.
The values of PC and SR are copied from the stack.
In \circled{4}, the memory space available on MSP430/MSP432 is limited, with memory sizes around 256KB.
So saving the register states of the entire IPE process on the target MSP microcontroller is not feasible and we use its UART interface to transmit the exfiltrated register state to a workstation for analysis.\footnote{We also experimentally verified that it is possible to use the debugger to transmit the exfiltrated register state. We found that using the UART was 22x faster than the debugger.}
In \circled{5}, the attacker's ISR resets the PC, SP, and SR registers to restore the program state to the attacker-controlled code.
The attacker code increments the timer counter value by one cycle and calls the victim code again.
We repeat phase 1 until the state-change due to all IPE code is captured.
The end result is similar to single-stepping through the IPE code.

\subsection{Post-processing (Phase 2)}
\label{design_phase2}

Phase 1 represents the real-time or dynamic part of the attack.
Phase 2 consists of post-processing the register state dumps and does not require access to the device.
The goal is to decode the IPE assembly, extracting as much information as possible from the register dumps.
\circled{6} Looking at the changes in PC, we deduce the instant an instruction is completed.
Comparing its PC with the PC of the next instruction, we are able to infer the size of the completed instruction unless it is some (taken) branch or call instruction, which is apparent if the PC jumps by a considerable value.
The frequency of successively repeating PCs gives the clock cycles required to complete the execution of that instruction.

It is important to note that in the MSP430 instruction set architecture, the instruction size and cycles depend only on the operands and not the operation.
An MSP430 instruction has broadly two types of operands, register or memory, with the possibility of combining these two for the source and destination operands. Identifying the operation is straightforward when both operands are direct register values.
In such cases, the critical piece of information is the change in values of the general-purpose registers (GPRs) and how those changes correlate to the previous values of the registers.
For instance, an \texttt{ADD} instruction with two register operands causes a difference in the value of the destination operand.
By doing so, we decode the operation as well as the operands.
However, we need to probe memory-to-register or register-to-memory instructions further in order to decipher them, as we detail in \S\ref{implement_phase2}.
We also observe that memory-to-memory operations are indecipherable just by looking at the register file, as at least one of the GPRs must change for identification.\footnote{The ARM instruction set of the MSP432 only allows for register-to-register state-changing operations, which increases the power of Phase 1 of our attack.} 

Status bit changes are also helpful in resolving the instruction guesses.
Changes in carry, zero, and negative bits clarify the state of the destination operand after the operation.
A carry bit set implies that the result of the operation produces a carry.
Zero or negative bit changes help us infer whether the result of an operation is zero or negative.
The overflow bit signifies an overflow in the signed-variable range~\cite{msp430cpux2012}.
It also helps resolve certain operations, e.g., an \texttt{AND} instruction resets the overflow bit.
These distinctions are essential when the destination operand is not a direct register value.

After reverse engineering, we are able to decode all the register and register-indirect addressing mode operations, where both operands are in one of those addressing modes.
\circled{7} The next step in post-processing depends on whether the attacker aims to obtain the key or exfiltrate the entire IPE memory.
If the former is true, the attacker needs to identify the function that operates on the key, such as the \texttt{KeyExpansion} function in AES.
We show in \S\ref{eval_main} that all the assemblies produced from popular cryptographic algorithms leak the key bits to the registers and phase 3 is not required.
However, if the attacker’s goal is to extract the entire IPE memory, we move on to the next phase.

\subsection{Exfiltrating IPE memory (Phase 3)}
\label{design_phase3}

An ROP gadget is a security exploit that allows an attacker to execute code in the presence of code protection mechanisms~\cite{checkoway2010return, roemer2012return}. 
For TI's IPE, there is no context switch gateway between calls from untrusted code to trusted code.
We observe that an attacker has the capability to branch/jump to any instruction inside the IPE code without breaching the IPE security.
Taking inspiration from ROP, we devise an interrupt- and data-oriented exploit for memory exfiltration purposes, which we call \textit{Interrupt-Oriented Programming} (IOP).
The fundamental idea of IOP is that if we setup the registers for a specific IPE instruction, branch directly to that IPE instruction, wait for it to complete execution, interrupt execution before the following instruction executes, and look at the results in the attacker's ISR, we get a Turing-complete set required for return-oriented programming~\cite{checkoway2010return, roemer2012return}.
Register indirect addressing mode instructions in the IPE code that directly modify the value of a destination register are ideal candidates for memory exfiltration as these instructions enable the attacker to read an IPE memory location by modifying the memory address value in the source register, followed by a readout of the value returned in the destination register.
Henceforth, we refer to this class of instructions as \textit{IOP gadgets}.

\circled{8} The first step of IP exfiltration is to find an IOP gadget.
\circled{9} Next, the attacker injects code into the non-IPE memory that sets up the timer module with a counter value, allowing the execution of the targeted instruction to complete.
The malicious function also writes the desired IPE segment base address to the source register of the instruction.
Then the program jumps to the IOP gadget inside the IPE zone.
After execution of the gadget instruction(s), the timer module interrupts IPE execution, invoking the attacker's ISR.
The ISR sends the value of the destination register to the attacker workstation and restores the PC, SP, and SR of the program to the attacker's control code.
Next, the malicious code increments the address location passed to the target instruction's source register depending on the operand's size.
Repeating this process, the attacker is able to get the dump of the desired IPE memory without causing any security violations.  

\section{RIPencapsulation Implementation}
\label{implement_main}

\begin{table}[!t]
    \centering
    \begin{tabular}{c c c}
        \toprule
        \small{\textbf{Platform}} & \small{\textbf{MSP430FR5994}} & \small{\textbf{MSP432P401R}}\\
        \midrule
        \small{ISA}                                 & \small{MSP430}        & \small{Armv7E-M}\\
        \small{NV Memory Type}                             & \small{FRAM}               & \small{Flash}\\
        \small{NV Memory Size}                             & \small{256 KB}             & \small{256 KB}\\
        \small{SRAM Size}                                    & \small{8 KB}               & \small{64 KB}\\
        \small{Max Clock Frequency}                              & \small{16 MHz}  & \small{48 MHz}\\
        \small{IP Encapsulation}                             & \small{Yes}                & \small{Yes}\\
        \small{UART Interface}                               & \small{Yes}                & \small{Yes}\\
        \small{Timer Module}                                 & \small{Yes}                & \small{Yes}\\
        \bottomrule
        \\
    \end{tabular}
    \caption{Microcontrollers we implement RIPencapsulation on.}
    \label{tab:msp430fr5994_specs}
\end{table}

We implement and evaluate RIPencapsulation on Texas Instruments MSP430FR5994 and MSP432P401R launchpads (commercially available development boards).
Table~\ref{tab:msp430fr5994_specs} details the relevant specifications of the respective microcontrollers. 
Since the MSP430 and MSP432 are based on different instruction set architectures, there are some implementation differences in RIPencapsulation on the two microcontrollers;
\S\ref{ripe_msp432} covers the implementation differences for the MSP432 and this section details the MSP430 implementation.

\subsection{Capturing IPE Register State}
\label{implement_phase1}

\begin{algorithm}[!t]
\caption{Interrupt-based side-channel routine for exfiltrating IPE process register states}\label{alg:phase1}
\begin{algorithmic}[1]
\STATE \emph{...Set timer ISR routine...}
\STATE $i \gets 0$
\STATE $TIMER\_COUNT \gets 0$
\WHILE{$i \neq DESIRED\_DUMPS$}
\STATE $TIMER\_COUNT \gets TIMER\_COUNT + 1$
\STATE \emph{...Set timer and UART parameters...}
\STATE \emph{...Start timer counter...}
\STATE \emph{...Enable interrupts...}
\STATE \emph{CALL IPE\_Function()}
\STATE // Secure process gets interrupted
\STATE \emph{...Copy all core registers to unprotected memory...}
\STATE \emph{...Copy saved state to UART TX Buffer...}
\STATE $i \gets i + 1$
\STATE \emph{...Reset SP, SR and PC registers...}
\ENDWHILE
\end{algorithmic}
\end{algorithm}

RIPencapsulation's interrupt-based side-channel procedure transfers the single-cycle register state dumps to the attacker's machine.
Algorithm~\ref{alg:phase1} provides the details of RIPencapsulation’s side channel routine.
MSP430FR5994 comes with two asynchronous general-purpose timer modules, each with four operating modes, and supports multiple captures/compares.\footnote{We use the timer module instead of the watchdog timer because it has finer-grain control over timer interval ranges and supports interrupt intervals as low as one clock cycle.}
The three operating modes besides \texttt{Stop} mode are – \texttt{Up}, \texttt{Continuous}, and \texttt{Up/down}; they have similar behavior:
the timer counts to a value before overflowing and restarting the count.
It generates interrupts when the timer counter overflows or reaches the \texttt{capture/compare} register value, depending on the operating mode.

After setting up the timer parameters, we start the timer and enable interrupts. Then we branch to the IPE code, where the timer interrupts it, invoking the interrupt service routine (ISR).
Inside the ISR, we copy the latent IPE state from the register file and sent it to the workstation via the UART.
Next, the ISR restores \texttt{PC}, \texttt{SP}, and \texttt{SR} to the attacker's state machine code, where we reset the timer parameters and restart the counter with the incremented \texttt{TIMER\_COUNT} value so that the device goes one cycle deeper into the IPE code.

\subsection{Reverse Engineering and Heuristic Analysis}
\label{implement_phase2}

\begin{figure*}[!t]
\centerline{\includegraphics[width=14cm, height=6cm]{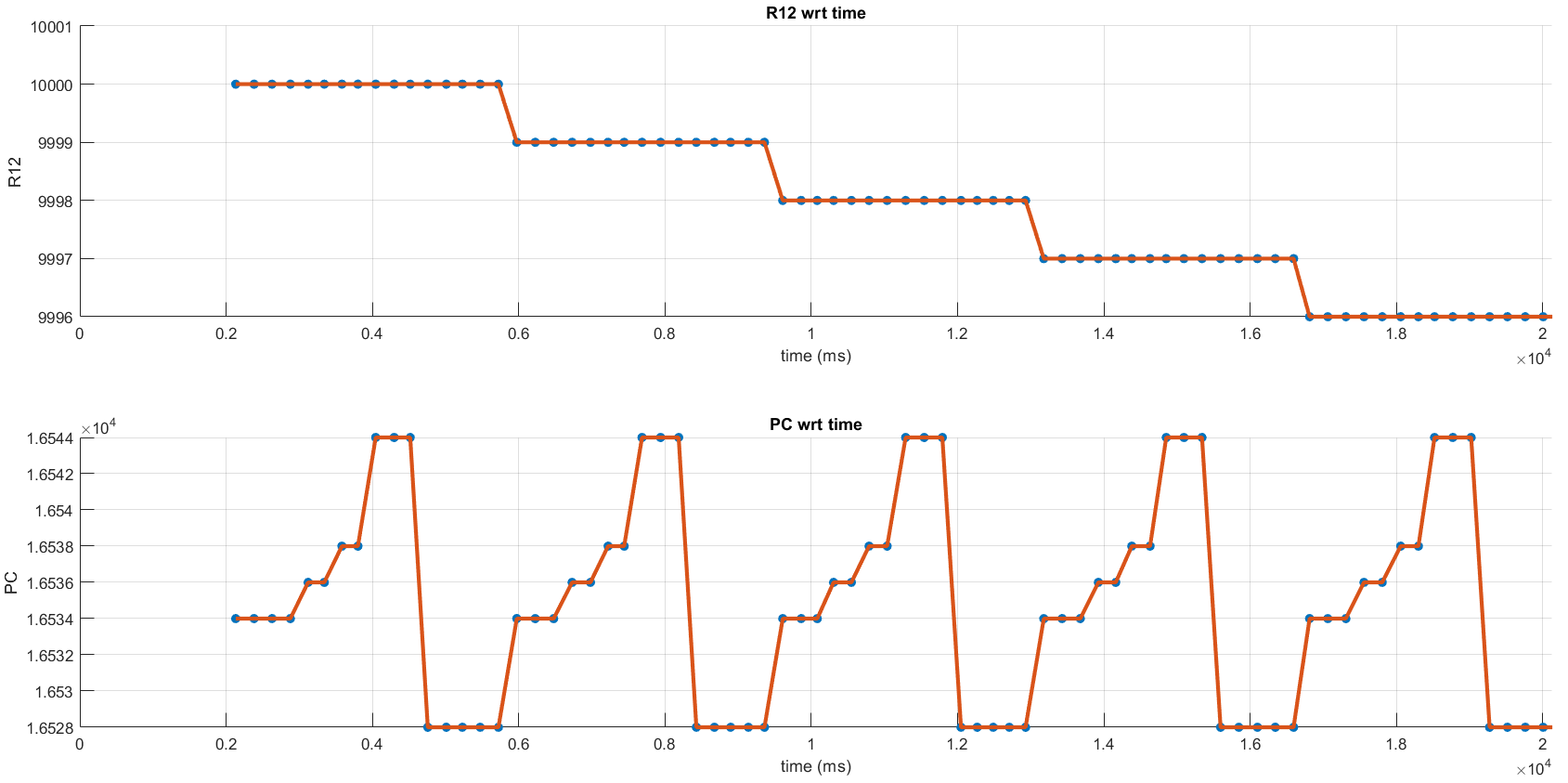}}
\caption{Register state plot of a simple decrementing loop-counter function showing the change of R12 register with the program counter.}
\label{simple_program_flowchart}
\end{figure*}

\begin{table}[htbp]
    \centering
    \begin{tabular}{@{} *{6}{c} @{}}
    \headercell{\textbf{Source}\\\textbf{Operand}} & \multicolumn{5}{c@{}}{\textbf{Destination Operand}}\\
    \cmidrule(l){2-6}
    & \textbf{Rm} & \textbf{PC} & \textbf{x(Rm)} & \textbf{TONI} & \textbf{\&TONI} \\ 
    \midrule
      \textbf{Rn}    & \checkmark &  \checkmark &  \checkmark &  $\times$ &  $\times$ \\
      \textbf{@Rn}   & \checkmark &  \checkmark &  \checkmark &  $\times$ &  $\times$ \\
      \textbf{@Rn+}  & \checkmark &  \checkmark &  \checkmark &  $\times$ &  $\times$ \\
      \textbf{\#N}   & -          &  -          &  -          &  $\times$ &  $\times$ \\
      \textbf{x(Rn)} & \checkmark &  \checkmark &  \checkmark &  $\times$ &  $\times$ \\
      \textbf{EDE}   & $\times$   &  $\times$   &  $\times$   &  $\times$ &  $\times$ \\
      \textbf{\&EDE} & $\times$   &  $\times$   &  $\times$   &  $\times$ &  $\times$ \\
      \bottomrule
      \\
    \end{tabular}
    \caption{MSP430 decoded instructions. EDE and TONI are generic labels and have no special meaning. Certain instructions cannot be guessed accurately as there is ambiguity whether the source operand is an immediate value or some label (-).}
    \label{tab:decoded_assembly}
\end{table}

MSP430 consist of the 27 implemented instructions of the MSP430 CPU~\cite{msp430cpux2012}. Besides this, some MSP430 devices also support additional \texttt{CPUX} instructions that are used when a greater than 16-bit address space is required.
The instructions are categorized into three types based on the operand sizes: byte, word, and 20-bit.
The MSP430 instructions support seven different addressing modes. 
Table~\ref{tab:decoded_assembly} summarizes the various source and destination addressing modes that the MSP430 instruction set architecture supports and highlights the instructions RIPencapsulation is able to guess correctly. RIPencapsulation successfully decodes the instructions that use register/indexed/indirect register mode source and destination operands.
It can also decipher the indirect autoincrement mode instructions, where the source operand is \texttt{@Rn+}.
In contrast to the indirect register mode instructions, these instructions increment the source register value by 1, 2, or 4 depending on whether the operand size is byte, word, or 20-bit respectively.
It is important to note that these four types of instructions constitute the bulk of the assembly in our benchmarks.
We are unable to decipher instructions with symbolic or absolute addressing mode source or destination operands as they are IP encapsulated and cannot be directly modified to attacker-controlled non-IPE address locations for disambiguation.
Also, we do not consider guesses for instructions with immediate mode source operand and register/indexed/indirect register mode destination operand in our evaluation of decoded assembly instructions, as it is ambiguous whether the source operand is an immediate value or a value stored in some IPE symbol or absolute address.

\begin{algorithm}[hbt!]
\caption{Sorting algorithm for register mode instructions}\label{alg:register_mode_sort}
\begin{algorithmic}[1]
\STATE $i \gets 0$
\WHILE{$i \neq TOTAL\_DUMPS - 1$}
\IF{$PC[i+1] - PC[i] \neq 0$}
\IF{$Rm[i+1] - Rm[i] \neq 0$}
\STATE $R_{dest} \gets Rm$ 
\ENDIF
\IF{$R_{dest}[i+1] \gets f(R_{dest}[i],Rn[i])$}
\STATE $R_{source} \gets Rn$
\STATE $OPERATION = f$
\ENDIF
\ENDIF
\STATE $i \gets i + 1$
\ENDWHILE
\end{algorithmic}
\end{algorithm}

Algorithm~\ref{alg:register_mode_sort} details the sorting algorithm for register mode instructions where the source and destination operands are register contents. 
For indexed or indirect register mode instructions, the source or destination operand contains a register value used as a pointer to the actual operand value. 
Algorithm~\ref{alg:indirect_mode_sort} highlights the procedure for identifying these instructions.
Before branching to the suspected instruction, the attacker sets all the general-purpose registers (GPRs) to controlled non-IPE memory addresses.
Based on the addressing mode, the suspected instruction will modify either the corresponding destination register or the memory address value it points to.
Observing which register or memory address value changes, we guess the destination operand and its addressing mode (register/indexed/indirect).
Analyzing the new value at the destination location and comparing it with the older GPR values, we are able to figure out the source operand, its addressing mode, and the operation performed.
For instance, a \texttt{"MOV @R10, R15"} loads the value at the memory address pointed by \texttt{R10} to \texttt{R15}.
For indexed instructions, we perform multiple tests with different memory address inputs to get a complete resolution, as we also need to identify the offset to the address value stored in the register operand.
Besides this, we also perform multiple tests to disambiguate unintended clashes between operations.
For instance, \texttt{XOR} between \texttt{0xFE} and \texttt{0x00} produces the same output as \texttt{MOV} with \texttt{0xFE}. 

\begin{algorithm}[!t]
\caption{Sorting algorithm for indexed and indirect mode instructions}\label{alg:indirect_mode_sort}
\begin{algorithmic}[1]
\STATE \emph{...Set timer ISR routine...}
\STATE $i \gets 0$
\STATE $TIMER\_COUNT \gets CYCLES\_REQUIRED$
\STATE $ADDR \gets SUSPECTED\_INSTRUCTION$
\WHILE{$i \neq NUMER\_OF\_TESTS$}
\STATE \emph{...Set timer and UART parameters...}
\STATE \emph{...Start timer counter...}
\STATE \emph{...Enable interrupts...}
\STATE \emph{...Set all GPRs to controlled memory addresses...}
\STATE \emph{JUMP ADDR}
\STATE // Secure process gets interrupted
\STATE \emph{...Copy all core registers to unprotected memory...}
\STATE \emph{...Copy saved state to UART TX Buffer...}
\STATE $i \gets i + 1$
\STATE \emph{...Reset SP, SR and PC registers...}
\ENDWHILE
\end{algorithmic}
\end{algorithm}

Single operand instructions such as \texttt{PUSH}, \texttt{POP}, \texttt{CALL}, and \texttt{RETI} modify the stack and have unique effects on the stack pointer, giving away their identity.
We also perform heuristic analysis on the obtained register dumps to clarify the victim program assembly state better.
Figure~\ref{simple_program_flowchart} illustrates the register state plot of a  decrementing loop-counter program.
Here the \texttt{R12} register holds the counter value decremented at the end of each loop.
The return of the program counter to the lower memory address signifies \texttt{JUMP}ing back to the start of the loop.
We interpret \texttt{PC} value discontinuities which do not modify the \texttt{SP} as \texttt{JUMP} instructions.

\subsection{IPE Memory Exfiltration}
\label{implement_phase3}

\begin{algorithm}[!t]
\caption{IP exfiltration routine using IOP gadget}\label{alg:phase3}
\begin{algorithmic}[1]
\STATE \emph{...Set timer ISR routine...}
\STATE $i \gets 0$
\STATE $TIMER\_COUNT \gets CYCLES\_REQUIRED$
\STATE $ADDR \gets IPE\_START - OPERAND\_SIZE$
\WHILE{$i \neq DESIRED\_DUMPS$}
\STATE \emph{...Set timer and UART parameters...}
\STATE \emph{...Start timer counter...}
\STATE \emph{...Enable interrupts...}
\STATE $ADDR \gets ADDR + OPERAND\_SIZE$
\STATE $R_{source} \gets ADDR$
\STATE \emph{JUMP IOP\_Gadget}
\STATE // Secure process gets interrupted
\STATE \emph{...Copy $R_{dest}$ value to unprotected memory...}
\STATE \emph{...Copy saved state to UART TX Buffer...}
\STATE $i \gets i + 1$
\STATE \emph{...Reset SP, SR and PC registers...}
\ENDWHILE
\end{algorithmic}
\end{algorithm}

Instructions which use indexed/indirect register mode source operand are ideal for exfiltrating the IPE memory.
However, any indexed/indirect register value also works as we modify the memory address pointed to by that register to some attacker-controlled memory location. 
After the attacker identifies an IOP gadget, they inject malicious code that writes the IPE segment base address to the source register of the victim instruction and branches to it.
Following the timer interrupt, the ISR transmits the saved destination operand containing the IPE memory value over the UART.
The malicious code then increments the victim address before the ISR transmits the next destination operand value to the attacker's machine.
Algorithm~\ref{alg:phase3} details RIPencapsulation's IP exfiltration routine using the IOP gadget.
\section{RIPencapsulation on MSP432}
\label{ripe_msp432}

We also test the efficacy of RIPencapsulation on MSP432 with the Cortex-M4 Flash-based microcontroller MSP432P401R.
This device has a different ISA, different system-on-chip, and different non-volatile memory, so by showing that RIPencapsulation works for the MSP432, we show that the flaws in TI's IPE are fundamental.

\subsection{CPU Halts Inside IPE Zone}
\label{msp432_1}

MSP432 IP Encapsulation has some key differences in its implementation, besides the fact that it is only configurable using the bootloader (due to Flash vs. FRAM differences). 
MSP432 IPE is more secure in that it disables all kinds of CPU halts inside the IPE zone. 
The \texttt{SYSCTL} security control monitors all debug accesses from the DAP (debug) port inside the IPE zone. 
In fact, \texttt{SYSCTL} does not permit any CPU halts inside the IPE zone, including breakpoint addresses pointing inside the IPE memory (unlike MSP430 IPE).
Even with this precaution, we find that we are able to trigger timer interrupts inside the IPE zone. 
Thus, like for the MSP430, we use timer interrupts to build our attack, however there are some differences in the interrupt handler routine. 

\subsection{Return from Interrupt Handler}
\label{msp432_2}

The setup procedure for timer interrupts is the same as on the MSP4320. 
However, the working of interrupts is slightly different. 
On the MSP432, all interrupts and exceptions are handled by the Nested Vectored Interrupt Controller (NVIC). 
When the NVIC detects that an interrupt signal is \texttt{HIGH}, it changes the state of the interrupt to pending. 
Interrupts remain pending until the processor enters the interrupt service routine (ISR), upon which the NVIC changes the interrupt status to active. 
When the interrupt is serviced by the ISR, the processor loads an \texttt{EXC\_RETURN} value stored in the link register (\texttt{LR}) to the program counter (\texttt{PC}). 
In our test code using timer interrupts, an \texttt{EXC\_RETURN} value of \texttt{0xFFFFFFE9} is loaded into \texttt{PC}, which returns to the original thread using the state from the main stack pointer (\texttt{MSP}). 
On return from the ISR, the NVIC changes the interrupt status from active to inactive. 
As such, when we try to replicate RIPencapsulation on the MSP432 by jumping to the malicious code, like we do on the MSP430, the interrupt status remains active, and the NVIC does not allow re-entry into the ISR even when the timer interrupt becomes pending again. 
To resolve this, we exit the ISR using the \texttt{EXC\_RETURN} value. 
So, instead of jumping directly to the malicious code, we write the malicious return state on the \texttt{MSP}.
This way, when the \texttt{EXC\_RETURN} value is loaded into the \texttt{PC}, the processor returns to the malicious code instead of the original thread, where we reset the timer interrupts and increment the timer counter for single-step execution of the victim IPE code. 

\subsection{Clearing Interrupt Flags}
\label{msp432_3}

Even though the program control returns to the malicious code upon completion of the ISR, as we desire, we observe that the program immediately re-enters the ISR, not executing any victim code. 
We discover that this is happening because the interrupt signal stays asserted even after completion of the ISR. 
The NVIC detects the asserted signal and immediately changes the interrupt status to pending, leading to re-entry inside the ISR. 
Clearing the interrupt flags de-asserts the interrupt signal and the NVIC waits for the next timer interrupt upon completion of the ISR.

\subsection{Unlocking Read Access}
\label{msp432_4}

Exfiltrating the data from the MSP432 IPE region is not so straightforward. 
The attacker must unlock data access each time the control goes inside the IPE zone. 
If we try to jump to an arbitrary load instruction inside the IPE zone without unlocking the data access first, the processor throws an exception. Writing \texttt{0x695A} to the memory-mapped \texttt{SYS\_SECDATA\_UNLOCK} register unlocks the data access for the IPE region writing to that register (since MSP432 allows the creation of up to four isolated IPE regions).
To bypass this, we need to construct a more sophisticated read IOP gadget.
In our proof-of-concept attack, we jump inside the IPE zone to a store instruction to write the unlock value to the memory-mapped \texttt{SYS\_SECDATA\_UNLOCK} register followed by an indirect register branch instruction to branch to the actual load instruction which we then use to exfiltrate the IPE firmware.
The attacker is capable of leveraging other sophisticated IOP gadgets to perform this read exploit.

\section{Evaluation}
\label{eval_main}

We evaluate RIPencapsulation against four commodity cryptographic algorithms, namely AES (tiny AES), SHA256 (saddi), SHA256 (gladman), and RSA (codebase).
We select these specific benchmarks in order to evaluate RIPencapsulation across different forms of cryptographic algorithms like symmetric-key and public-key cryptography and cryptographic hashing.
We compile all benchmarks using the open-source MSP430 GCC toolchain developed by Texas Instruments~\cite{ti:slau646f}, testing them against a range of optimization levels.
Our evaluation answers the following questions:
\begin{itemize}
    \item Are cryptographic implementations generally susceptible to the RIPencapsulation attack?
    \item What effect does the compiler optimization level have on the vulnerability of these cryptographic implementations?
    \item How long does it take to carry out the RIPencapsulation attack?
    \item Is IPE vulnerable across device types?
\end{itemize}

\subsection{AES}

\begin{table}[!t]
    \centering
    \begin{tabular}{c c c c}
        \toprule
        \small{\textbf{Optimization}} & \small{\textbf{Instructions}} & \small{\textbf{Reveal key}}  & \small{\textbf{Contains IOP}}\\
        \small{\textbf{Level}}              & \small{\textbf{decoded}}      & \small{\textbf{bits?}}     & \small{\textbf{Gadget?}}\\
        \midrule
        -O0                          &60.5\%                 & \checkmark                & \checkmark\\
        -Og                          &66.6\%                 & \checkmark                & \checkmark\\
        -O1                          &68.4\%                 & \checkmark                & \checkmark\\
        -O2                          &68.4\%                 & \checkmark                & \checkmark\\
        -O3                          &68.4\%                 & \checkmark                & \checkmark\\
        -Os                          &67.9\%                 & \checkmark                & \checkmark\\
        -Ofast                       &68.4\%                 & \checkmark                & \checkmark\\
        \bottomrule
        \\
    \end{tabular}
    \caption{Test cases with different compiler optimizations for AES (tiny AES). The second column depicts the percentage of assembly instructions we are able to reverse engineer after Phase 2.}
    \label{tab:aes_result}
\end{table}

Our static analysis results show that the AES implementation leaks key bits to the registers in its \texttt{KeyExpansion} function. At optimization levels \texttt{O1} and above, the \texttt{KeyExpansion} function is embedded inside the \texttt{AES\_init\_ctx} function. All the instructions leaking the key bits are of the form \texttt{MOV.B @Rn, Rm}. The destination registers in these instructions leak the last four bytes of the secret key and all the remaining round keys. Reverse engineering the AES secret key using the round keys is a deterministic process, and literature exists describing the same~\cite{cordwell2008aes}. We take the second-round key and the last four bytes of the secret key to obtain the original 128-bit key using the following formula:

\begin{quote}
    \textit{If} $i\%4\neq0,$
\end{quote}
\begin{displaymath}
    \omega_{i-4} = \omega_{i}  \oplus (\omega_{i-1})
\end{displaymath}
\begin{quote}
    \textit{Else,}
\end{quote}
\begin{displaymath}
    \omega_{i-4} = \omega_{i}  \oplus sbox(shift(\omega_{i-1} )),
\end{displaymath}
\begin{quote}
    \textit{followed by XOR of} $\omega_{i-4}$\textit{'s 1st byte with} $Rcon[j]$ \\
\end{quote}

Here $\omega_{i}$ is the $ith$ word in the complete AES key. $sbox(x)$ represents the byte substitution using the S-Box lookup table, $shift(x)$ is a cyclical shift of the bytes of $\omega_{i}$, and $Rcon[j]$ is the round constant for round $j$, whose key we reverse engineer here. This would be the first round in our case. It is also worth mentioning that the \texttt{-O0} optimization level assembly reveals the secret key location in the stack. Assemblies produced at all optimization levels contain an instruction of the form \texttt{MOV.B @Rn, Rm}. We use them as IOP gadgets to exfiltrate the AES IPE code and data. Table~\ref{tab:aes_result} summarizes the results for our AES-128 implementation.

\subsection{SHA256}

\begin{table}[htbp]
    \centering
    \begin{tabular}{c c c c}
        \toprule
        \small{\textbf{Optimization}} & \small{\textbf{Instructions}} & \small{\textbf{Reveal `pt'}}  & \small{\textbf{Contains IOP}}\\
        \small{\textbf{Level}}              & \small{\textbf{decoded}}      & \small{\textbf{bits?}}     & \small{\textbf{Gadget?}}\\
        \midrule
        -O0                          &59.4\%                 & \checkmark                & \checkmark\\
        -Og                          &56.7\%                 & \checkmark                & \checkmark\\
        -O1                          &53.9\%                 & \checkmark                & \checkmark\\
        -O2                          &49.7\%                 & \checkmark                & \checkmark\\
        -O3                          &50.1\%                 & \checkmark                & \checkmark\\
        -Os                          &53\%                 & \checkmark                & \checkmark\\
        -Ofast                       &50.1\%                 & \checkmark                & \checkmark\\
        \bottomrule
        \\
    \end{tabular}
    \caption{Test cases with different compiler optimizations for SHA256 (saddi). The second column depicts the percentage of assembly instructions we are able to reverse engineer after Phase 2.}
    \label{tab:sha_saddi_result}
\end{table}

Table~\ref{tab:sha_saddi_result} highlights the results of the RIPencapsulation attack on saddi SHA256 implementation. Not only do all the assemblies leak the plaintext location to the registers, but also all the plaintext bits. The plaintext bits are leaked in the \texttt{SHA256Guts} function, which is embedded inside the \texttt{sha256\_update} function at optimization levels \texttt{-O1} and above. The instructions containing the leaking registers are word instructions, and the plaintext bytes are present in reverse order within each word as the memory model that MSP430FR5994 uses is little endian. All our SHA256 assemblies contain instructions of the form \texttt{MOV.W x(Rn), Rm}. We use them as IOP gadgets to exfiltrate our SHA256 IPE code and data.

\begin{table}[htbp]
    \centering
    \begin{tabular}{c c c c}
        \toprule
        \small{\textbf{Optimization}} & \small{\textbf{Instructions}} & \small{\textbf{Reveal `pt'}}  & \small{\textbf{Contains IOP}}\\
        \small{\textbf{Level}}              & \small{\textbf{decoded}}      & \small{\textbf{bits?}}     & \small{\textbf{Gadget?}}\\
        \midrule
        -O0                          &59.2\%                 & \checkmark                & \checkmark\\
        -Og                          &60.9\%                 & \checkmark                & \checkmark\\
        -O1                          &58.3\%                 & \checkmark                & \checkmark\\
        -O2                          &60.4\%                 & \checkmark                & \checkmark\\
        -O3                          &60.7\%                 & \checkmark                & \checkmark\\
        -Os                          &57.7\%                 & \checkmark                & \checkmark\\
        -Ofast                       &60.7\%                 & \checkmark                & \checkmark\\
        \bottomrule
        \\
    \end{tabular}
    \caption{Test cases with different compiler optimizations for SHA256 (gladman). The second column depicts the percentage of assembly instructions we are able to reverse engineer after Phase 2.}
    \label{tab:sha_gladman_result}
\end{table}

We also evaluate RIPencapsulation on the gladman SHA256 implementation. Table~\ref{tab:sha_gladman_result} summarizes its evaluation results. All assemblies of this SHA256 implementation leak the plaintext bits to the registers inside the \texttt{sha\_end1} function. The compiled code also contains a prologue instruction inside the calling function for SHA256, which performs a \texttt{SUBA} operation on the stack pointer. We bypass this instruction and, as such, include it in our calling routine for the IP-encapsulated SHA256 function. Assemblies produced at all optimization levels contain IOP gadgets that look like \texttt{MOV.W @Rn, Rm}.

\subsection{RSA}

\begin{table}[htbp]
    \centering
    \begin{tabular}{c c c c}
        \toprule
        \small{\textbf{Optimization}} & \small{\textbf{Instructions}} & \small{\textbf{Reveal key}}  & \small{\textbf{Contains IOP}}\\
        \small{\textbf{Level}}              & \small{\textbf{decoded}}      & \small{\textbf{bits?}}     & \small{\textbf{Gadget?}}\\
        \midrule
        -O0                          &58.3\%                 & \checkmark                & \checkmark\\
        -Og                          &58.4\%                 & \checkmark                & \checkmark\\
        -O1                          &57.3\%                 & \checkmark                & \checkmark\\
        -O2                          &54.3\%                 & \checkmark                & \checkmark\\
        -O3                          &54.3\%                 & \checkmark                & \checkmark\\
        -Os                          &60.8\%                 & \hspace{1pt} \checkmark*  & \checkmark\\
        -Ofast                       &54.3\%                 & \checkmark                & \checkmark\\
        \bottomrule
        \\
    \end{tabular}
    \caption{Test cases with different compiler optimizations for RSA (codebase). The second column depicts the percentage of assembly instructions we are able to reverse engineer after Phase Two. *Readout from the stack is required for some bits of the private key.}
    \label{tab:rsa_result}
\end{table}

The codebase RSA implementation is the smallest code of all the benchmarks evaluated in this paper. Table~\ref{tab:rsa_result} summarizes the evaluation results for our RSA implementation. Assemblies produced at optimization levels \texttt{-O0}, \texttt{-Og}, and \texttt{-O1} store the private key location in the stack and leak the private key bits to the registers inside the \texttt{modexp} function, which is called by the \texttt{rsaDecrypt} function. Level \texttt{-O2}, \texttt{-O3}, and \texttt{-Ofast} leak both the private key location and the key itself to the general-purpose registers inside the \texttt{rsaDecrypt} function. The \texttt{-Os} assembly stores the private key location in the stack and only partially leaks the private key bits to the registers inside the \texttt{modexp} function. The assembly directly leaks the first two words to the registers but accesses the next two words in the \texttt{BIS.W x(SP), R12} instruction, which performs an \texttt{OR} between the two operands. Since \texttt{OR} is lossy, we cannot reverse-engineer the operands from the result of the operation. So, we need to read out the last two words of the private key from the stack. The \texttt{-Os} optimization assembly contains an IOP gadget of the form \texttt{MOV.W @SP, R12}, that is used to exfiltrate the private key bits from the stack. The \texttt{-O0} level produces assembly with a prologue consisting of a \texttt{PUSH} operation on the stack followed by a \texttt{SUBA} operation on the stack pointer. All other assemblies of RSA contain only the \texttt{SUBA} operation on the stack pointer in the calling function.

\subsection{Attack Time}
IPE exfiltration rate is characterized by the delay in transmitting the data packets to the external device over the UART communication channel. At a baud rate of 115200 and clock frequency of 1 MHz, it takes 10 seconds on average to obtain the register states leaking the second round key of the tiny AES encryption process. Saddi SHA256 directly leaks the plaintext bits to the register states in under 75 seconds for the \texttt{O0} optimization level, whereas \texttt{Ofast} optimized code leaks the plaintext in 23 seconds. Although the entire gladman SHA256 register state exfiltration process takes the longest time due to its large code size (2x in size compared to saddi SHA256 and 4x compared to codebase RSA), it leaks the plaintext bits to the registers in roughly the same amount of time as saddi, ranging from 30 to 70 seconds, with the \texttt{O0} level assembly being the most time-consuming. RSA codebase is the simplest benchmark of all and leaks the private key to the registers within 5 seconds. However, since \texttt{Os} optimized RSA code requires exfiltration of some key bits from the stack, phase three is necessary and the attacker must use an IOP gadget for exfiltrating the private key. In general, we are able to receive 10,000 dumps over the UART channel in under 2 minutes. The time taken to execute phase two of the attack depends on the attacker's approach to analyzing the exfiltrated register state. To accelerate phase two, we automate the reverse engineering process on our workstation.

\subsection{MSP432 Results}
\label{msp432_res}

We were able to replicate all of the attacks from the MSP430 on the MSP432.
Even though the MSP432 is a newer ISA with a more security-oriented system-on-chip, the RIPencapsulation is easier on the MSP432 due to its Thumb architecture set, which predominantly uses register mode operations.
Thus, the reverse engineering capability of RIPencapsulation on MSP432 increases to 80\% without Phase 3. 

\subsection{Write Exploit}
\label{write_exploit}

Besides leaking the IPE memory to the outside world, specific IOP gadgets are also able to write to the IPE region, giving the attacker the power to modify the IPE code and possibly get access control over it that way. This breaks the integrity and authenticity guarantees of code and data inside the IPE region. Register direct/indirect addressing mode instructions in the IPE code with register indirect destination addressing are ideal candidates for such an exploit as they allow deploying a payload to any desired IPE memory location. For the MSP430 instruction set these instructions look like \texttt{MOV Rn, x(Rm)}, \texttt{MOV @Rn, x(Rm)}, \texttt{MOV x(Rn), x(Rm)}, \texttt{MOV Rn, x(SP)} and many more variations of the same. We find that all our MSP430 benchmarks contain one of these instructions across all compiler optimizations. Meanwhile, the MSP432 uses Flash for its non-volatile memory hence we need the Flash Controller for writes. However, MSP432 IPE only allows storing code and constants in its IPE region and the Flash Controller is disabled inside the IPE zone. 
\section{Mitigation Strategies}
\label{mitigation}

We find two fundamental shortcomings in Texas Instruments's (TI's) IP Encapsulation (IPE) design, which make it vulnerable to RIPencapsulation.
First, it does not clear the IPE state on context switches to non-IPE execution.
This enables the attacker to invoke an Interrupt Service Routine (ISR) outside the IPE zone but retain access to the latent state of IPE execution through the register file.
Secondly, IPE has no call site verification.
This allows the attacker to jump anywhere inside the IPE zone, opening the door to control- and data-oriented attacks.
Even though TI recommends the IP author to disable interrupts from non-IPE code as a secure coding practice, we show that this is futile as attackers can bypass any protections in the IPE zone by jumping to the instruction right after.
Thus, a defense must address both the fundamental IPE flaws.

``The privacy of register and on-chip caches should be protected by the trusted computing base from software attacks"~\cite{suh2003aegis}.
In the AEGIS Architecture~\cite{suh2003aegis}, a secure context manager (SCM) stores all the process's register values in the SCM table on interrupt and clears the register states before invoking the ISR so that the ISR cannot access the internal state of the secure process.
The SCM restores the register states from the SCM table on return from the ISR. Clercq et al.~\cite{de2014secure} provide a hybrid implementation of AEGIS for MSP430 devices that improves SANCUS~\cite{noorman2013sancus}.
When extended to include clearing the residual state present in the shared memory region of SRAM (e.g., the stack), AEGIS-based system should address the problem of latent IPE state upon unexpected IPE exits---at the cost of hardware modification.

The lack of valid call site verification allows the attacker to orchestrate data-oriented control flow attacks.
Almost 90\% of exploit-based software attacks use some form of Return-Oriented Programming (ROP)~\cite{Jarrous2016-ps}.
Address Space Layout Randomization (ASLR) is a well-known class of code security techniques~\cite{forrest1997building, xu2003transparent, kil2006address, iyer2010preventing, williams2016shuffler} that randomize the memory address of a program's sections in order to reduce the chances of code reuse exploits that rely on knowing the exact location of process objects.
Not only does ASLR entail a heavy overhead, but it also supposes a higher level of ability to intrude into the code and introspect its insecure uses.
The facilities available on MSP430 and MSP432 devices do not lend themselves to an efficient implementation of such a high overhead approach.

A better defense for control- and data-oriented attacks is a call gateway veneer.
In the embedded space, ARM TustZone enforces this gateway veneer by introducing a Non-Secure Callable (NSC) memory region.
All calls from a normal program to a secure function must go through the gateway veneer residing in the NSC memory.
Calls to invalid entry points inside the secure code cause a hardware exception which always traps into a secure state.
Fault injection~\cite{tang2017clkscrew} and short-term data remanence~\cite{mahmod2022sram} attacks break the security guarantees of ARM TrustZone by pausing the trusted execution in a controlled manner to reveal its internal state.

We believe that ARM TrustZone, combined with fault attack protection, in conjunction with AEGIS-style secure context-switching represents the best solution to defend IoT-class devices against RIPencapsulation.
Given the many tradeoffs at play for ultra-constrained devices like MSPs, we believe that the design, implementation, and evaluation of such a defense is important future work that this paper motivates.

\section{Related Work}

Trusted Execution Environments (TEEs) are process isolation and secure storage solutions that are finding their way into IoT-class embedded devices.
Recent work indicates a rise in the trend of secure process preemption or exception-based exploits, which infer the program's internal state by studying their effects on the unprotected areas of the compromised device.
In light of this trend and given RIPencapsulation is a TEE attack, we cover attacks against other TEEs on devices ranging from desktop-class to embedded devices.

\subsection{Interrupt-based Attacks}

SGX-Step \cite{van2017sgx} is an interrupt-based side-channel attack on Intel Software Guard eXtensions (SGX), which builds on previous kernel-level SGX exploits that preempt the enclave execution to leak information from page tables (PTE)~\cite{xu2015controlled, van2017telling} or branch prediction units~\cite{lee2017inferring}. 
SGX-Step exploits the Advanced Programmable Interrupt Controller (APIC) timer to interrupt the secure process at several-instruction granularity in order to gain fine-grained control of the side channels, improving the temporal resolution of previous enclave preemption attacks~\cite{xu2015controlled, van2017telling, lee2017inferring, hahnel2017high}.
Nemesis~\cite{van2018nemesis} extends the SGX-Step by using an interrupt-based side channel to leak instruction-level information from TEE execution.
Nemesis requires precisely timed interrupts to capture per-instruction latency differences.

RIPencapsulation extends the idea of SGX-Step and Nemesis to IoT class devices.
RIPencapsulation provides single-cycle granularity and uncovers 100\% of TEE-protected instructions by combining interrupt-based attacks with data-oriented attack patterns. 
In additon, the compiler-based defenses that apply to heavyweight TEEs that rely on detecting high rates of page faults or interrupts by leveraging the x86 Transitional Synchronization eXtensions (TSX) do not apply in IoT-class devices~\cite{shih2017t, chen2017detecting}.


\subsection{Debugger-based Attacks}

Shedding too much Light on a Microcontroller's Firmware Protection~\cite{obermaier2017shedding} analyzes the security of protected Flash memory on STM32 microcontrollers. 
The paper uses fine-grain CPU resets and a vulnerability in the Flash protection logic protocol to extract the entire firmware by accessing iterative addresses of the firmware via the debugger and capturing latent Flash data in unprotected SRAM.
While this attack requires optical fault injection and chemical etching to manipulate particular Flash bits, RIPencapsulation can leverage their controlled CPU resets in the event that timer interrupts are not available.

Brosch~\cite{brosch2015firmware} presents a firmware dumping technique for an ARM Cortex-M0 SoC that uses the debugger to manipulate the CPU register values at single-step intervals. 
By single-stepping through the program code and observing the CPU register changes, they find load instructions to exfiltrate the protected memory. 
Their attack is based on the debugger's single-stepping capability inside protected memory, which is prevented by TI's IPE implementation.
RIPencapsulation side-steps TI's protection through a combination of interrupts and data-oriented attacks, achieving single-step IPE execution.

\subsection{Blind Attacks}

Code reuse attacks require varying degrees of information on the target. 
Blind attacks aim to create exploits when neither the source nor binary code is available.
Half-Blind attacks~\cite{goodspeed2009half} presents a stack overflow/ROP gadget attack to gain privileged access in the bootstrap loader (BSL) of an MSP430 device, which can then be used to extract the firmware image. 
Hacking Blind~\cite{bittau2014hacking} is a more generic and advanced version of half-blind attack which is fully blind and presents techniques to find and chain multiple, different gadgets. 
While these attacks do not work with TI's IPE protection, they serve as inspiration for attacking unknown binaries with RIPencapsulation.

\subsection{Other attacks on TEEs}

CipherLeaks~\cite{li2021cipherleaks} is a ciphertext side-channel attack on AMD's Secure Encrypted Virtualization (SEV) TEE.
SEV protects Virtual Machines (VM) from an untrusted hypervisor by using hardware-enforced memory encryption. 
On domain switch between the guest and host VM, SEV stores the encrypted register states in the Virtual Machine Save Area (VMSA).
The CipherLeaks attack model assumes that the attacker has read access to the VMSA but no write access. 
CipherLeaks uses the hypervisor to monitor specific offsets of the VMSA to infer changes of any 16-bit plaintext.
Non-Automatic VM Exits (NAE) expose some plaintext register values to the hypervisor. In essence, the attacker triggers an NAE to collect a dictionary of plaintext-ciphertext pairs for these registers stored in the VMSA.
Cipherleaks then uses this plaintext-ciphertext dictionary to crack the entire OpenSSL RSA key in 410 rounds with 100\% accuracy.
In order to patch this vulnerability, AMD added randomization in stored register values when encrypting and saving them into the VMSA during VMEXITs~\cite{amd2021sevpatch}.
This fix is available in the AMD SEV-SNP TEE.
Unfortunately, encrypting the register file is not a complete defense against RIPencapsulation, as the ability to enter the IPE zone at arbitrary points allows the attack to create IPE zone altering and exfiltration gadgets.

CLKscrew~\cite{tang2017clkscrew} is an energy management-based exploit that manipulates the voltage and frequency of the processor to induce faults.
Dynamic Voltage and Frequency Scaling (DVFS) is an energy management scheme, ubiquitous on commodity devices, that trades off processing speed for energy savings. 
CLKscrew is able to break the confidentiality and integrity of ARM TrustZone using software-only control of the regulators.
In essence, the attacker increases the frequency of the processor beyond the limits dictated by the operating voltage to induce instability and halt the TrustZone process from the normal world.
Performing Differential Fault Analysis on the correct and faulty decrypted plaintext pair, the attacker is able to infer the AES key.
We envision an exploit that combines the Interrupt-Oriented Programming described in this paper with CLKScrew to get very fine-grained control over fault injections inside the TEE.

\section{Conclusion}
Texas Instruments MSP IP Encapsulation (IPE) aims to provide confidentiality of data stored inside the IP-encapsulated memory zone;
this includes proprietary code and keys.
RIPencapsulation breaks this guarantee by leveraging two fundamental drawbacks in MSP430 IPE design: residual state on context switches and lack of call site verification.
We exploit these flaws to create an interrupt-based side channel to gain cycle-accurate control IPE execution and exfiltrate all IPE secrets.
The evaluation shows that this attack works using production tools and settings of popular open-source cryptographic implementations.

This paper shows that Trusted Execution Environment (TEE) designers must pay careful attention to unexpected TEE entries and exits.
Without guarding the entries to IPE code, attackers can bypass defenses and create gadget-like instruction sequences.
Without cleaning up residual state shared across security domains on every possible exit, attackers have access to secret-revealing side-channel information.
These requirements extend beyond any single TEE implementation, serving as necessary conditions for ensuring code and data confidentiality by all TEEs.

\section*{Acknowledgements}
The project depicted is sponsored by the Defense Advanced Research Projects Agency.
The content of the information does not necessarily reflect the position or the policy of the Government, and no official endorsement should be inferred.
Approved for public release; distribution is unlimited.

\bibliographystyle{plain}
\bibliography{references}

\end{document}